\documentclass[
 amsmath,amssymb,
 aps,
 prl,
 twocolumn,
 nofootinbib]{revtex4-2}

\usepackage{graphicx}
\usepackage{dcolumn}
\usepackage{bm}
\usepackage[normalem]{ulem} 
\usepackage[colorlinks=true,citecolor=blue,urlcolor=blue]{hyperref}
\usepackage{physics}
\usepackage{tensor}
\usepackage{xcolor}
\usepackage{comment}

\newcommand{\Austin}{\affiliation{Weinberg Institute, University of Texas at Austin, Austin, Texas 78712, USA}}

\newcommand{\varx}{\chi}
\newcommand{\kfactor}{\psi}
\newcommand{\lfactor}{\varphi}

\begin{document}

\title{New Framework for Classical Double Copies}

\author{Brian Kent and Aaron Zimmerman}
\Austin

\date{\today}

\begin{abstract}
The double copy relates gauge and gravitational theories, with widespread application to quantum scattering amplitudes and classical perturbative results.
It also connects exact classical solutions of Abelian gauge and gravitational theories in a small number of specific examples, known as classical double copies.
These have a number of special properties, such as being algebraically special, and it remains an open question whether examples exist for algebraically general spacetimes or with nontrivial dynamics.
Here we provide a novel framework for understanding classical double copies at the level of the metric, both organizing the known examples and exploring their properties under a new lens. 
Utilizing Killing vectors as natural gauge fields on a spacetime, we propose a procedure for generating new classical double copies. 
As a proof of concept, we provide a flat-space single copy for the Kasner metric, an algebraically general spacetime. 
This example is also a double copy at the level of its curvature, and provides the first type-$I$ Weyl double copy on flat spacetime, confirming expectations from perturbation theory. 
This provides a promising avenue for extending exact double copies to a broader class of general, physically relevant spacetimes.
\end{abstract}

\maketitle

\noindent \emph{Introduction.}---The \emph{classical double copy} is a connection between gauge fields in Minkowski space 
[or, more generally, a curved base spacetime, such as (anti--)de Sitter space] 
and solutions of Einstein's equations.
It provides a classical analog to the quantum field theory \emph{double copy} \cite{Bern:2008qj,Bern:2019prr}, which allows the computation of scattering amplitudes in gravitational theories from amplitudes in gauge theories. 
This is due to the fact that gauge theory amplitudes may be perturbatively reorganized into a product of color factors and kinematic factors which follow the same Lie-algebra identities as the color factors.
This color-kinematics duality makes it possible to replace color with kinematic information and vice versa. 

The most well-studied double copy relates pure Yang-Mills theory to a gravitational theory containing a graviton, two-form, and a dilaton.
The Yang-Mills amplitudes are themselves constructable from a biadjoint scalar theory \cite{Bern:2008qj,Bern:2010ue,Bern:2010yg}.
However, many more examples exist within the literature~\cite{Bern:2019prr}; e.g.,
widespread connections have been found in supergravity \cite{chiodaroli2017spontaneouslybrokenyangmillseinsteinsupergravities,Chiodaroli_2018} and string theory \cite{Stieberger_2016,Schlotterer_2016} and even to perturbative predictions of gravitational waves from compact binaries~\cite{Cheung_2018,Bern_2019}.
At tree level, these relations are low-energy limits of the Kawai-Lewellen-Tye relations \cite{Kawai:1985xq}, which relate open and closed strings. 
They have been shown to hold up to high loop orders in many examples, e.g.,~\cite{Bern:2010ue,Bern:2012uf,Bern:2019prr}, meaning such techniques have seen great utility for high perturbative order gravitational scattering in post-Minkowskian expansions \cite{Bern_2019,Kosower_2019,Bern:2022jvn}. This has led to speculation that double-copy relations may be present within exact solutions to the Einstein field equations.

The first example of an exact double-copy solution utilized 
Kerr-Schild spacetimes \cite{1965Kerr, gurses_lorentz_1975}
\begin{equation}
    g_{\mu \nu} = \eta_{\mu \nu} + \psi k_\mu k_\nu, 
    \label{KSDC}
\end{equation}
with $\eta_{\mu\nu}$ the flat metric and $k_\mu$ null and geodesic, 
mimicking the structure of a position-space double copy \cite{Monteiro:2014cda}. 
Specifically, considering $k_\mu$ as a kinematic factor and replacing $k_\mu \to c^a$, with $c^a$ a color factor, yields a gauge field called the single copy, $A^a_\nu = \psi k_\nu c^a$ obeying Abelian (linearized) Yang-Mills equations.
The ``graviton" is then $h_{\mu \nu} = \psi k_\mu k_\nu$ and the biadjoint scalar $\Phi^{ab} = \psi c^a c^b$, called the zeroth copy, satisfies Abelianized equations of motion.
Since these equations of motion are equivalent to flat-space Maxwell's equations for $A_\mu$ and wave equation for $\psi$, the color factors are generally neglected in the discussion of classical double copies.
This structure, visible at the level of the metric, is the \emph{Kerr-Schild double copy}~\cite{Monteiro:2014cda}.
Kerr-Schild spacetimes include the Kerr solution for a rotating black hole, and as such have wide-ranging applications in theoretical physics and astrophysics.

Since the original observation of this classical double copy, a large number of exact spacetimes with double- or single-copy structures have been identified, e.g.,~\cite{Monteiro:2014cda,Bahjat-Abbas:2017htu,CarrilloGonzalez:2019gof,Alkac:2021bav,Alkac:2021seh,Ortaggio1,Ortaggio2024,Bahjat-Abbas:2020cyb,Luna:2015paa,Luna:2016due,Luna:2018dpt,Chawla:2022ogv,Carrillo-Gonzalez:2017iyj,Gurses:2018ckx, Barrientos:2024uuq, Hassaine:2024mfs,LopesCardoso:2024ttc,Easson:2020esh,Godazgar:2020zbv,Easson:2021asd,Easson:2022zoh}.
In many cases, the double copy is visible at the level of the metric, with decompositions of the form of Eq.~\eqref{KSDC} or suitable generalizations \cite{Luna:2015paa,Luna:2018dpt}.
Closely related to this metric double copy is the Weyl double copy~\cite{Luna:2018dpt}, where curvature quantities are related to products of gauge field strengths and scalar fields \cite{Easson:2021asd,Easson:2022zoh,armstrongwilliams2024derivingweyldoublecopies,Chacon:2021wbr,Alkac:2023glx,Kent_2025,cdc-gem,Zhao:2024ljb,Godazgar:2021iae,Godazgar:2020zbv,Luna:2022dxo,Han:2022ubu}.
The Weyl double copy is generally studied in spinor language and phrased in terms of the Weyl spinor $\Psi_{ABCD}$, which for Petrov types $D$ and $N$ decomposes as 
\begin{equation}
    \Psi_{ABCD} = \frac{f_{(AB}f_{CD)}}{S}\,.
    \label{WDC}
\end{equation}
Here $f_{AB}$ is the spinorial analog of a two-form, $S$ is a (complex) scalar field, and in the simplest cases $f_{AB}$ obeys the flat-space Maxwell's equations and $S$ the flat-spacetime wave equation. 
However, the connection between the Weyl double copy and the metric double copy is still uncertain, and there exist
spacetimes with a Weyl double copy but for which any corresponding metric double copy  $\hat{g}_{\mu\nu}$ cannot be flat~\cite{Godazgar:2020zbv}.

Direct connections between three-point amplitudes in the perturbative double copy and the exact classical double copy have been identified  \cite{Arkani-Hamed:2019ymq,Emond:2020lwi,Huang:2019cja,Monteiro:2021ztt,Monteiro:2020plf,crawley2022black}.
These explain why the classical double copies identified thus far satisfy Abelian Yang-Mills; nonlinearities do not factor into three-point amplitudes between a point charge and a gluon.
Furthermore, a correspondence between the perturbative double copy and the Weyl double copy has been realized at the linear level through twistor theory~\cite{White:2020sfn,Chacon:2021lox,Chacon:2021wbr,Luna:2022dxo,guevara2021reconstructing,Adamo:2021dfg}.
Specifically, three-point amplitudes in momentum space can be converted via the Penrose transform into position space, resulting in a splitting of the Weyl spinor as in Eq.~\eqref{WDC} for Petrov type-$D$ spacetimes~\cite{guevara2021reconstructing,Luna:2022dxo}. 
Twistor methods have also been used to show that the linearized Weyl double copies of algebraically general spacetimes (Petrov type I) have the form
\begin{equation}
   \Psi_{ABCD} = \sum_{ij} \frac{f^{(i)}_{(AB}f^{(j)}_{CD)}}{S^{(ij)}}, \label{twistorweyl}
\end{equation}
where $f^{(i)}_{AB}$ are distinct electromagnetic spinors and $S^{(ij)}$ distinct scalars~\cite{Chacon:2021wbr}.
However, to date, no exact algebraically general classical double copies have been found. 
Furthermore, efforts to build spacetimes from the perturbative double copy order by order yield~\cite{Luna_2017,Kim:2019jwm} dilatonic gravity spacetimes~\cite{JNW} which are algebraically general. 
If generic metric double copies exist, these too should be algebraically general.

Despite extensive progress in the field of classical double-copy solutions, many questions remain.
At the most basic level, there is no sufficiently general definition of the metric double copy.
In this Letter we identify common features among vacuum spacetimes with known metric double copies, which, in turn, suggest avenues for seeking double copies of greater generality.
Armed with these insights we identify new metric double copies. 
We show that a known class of generalized Kerr-Schild spacetimes possess a classical double copy. 
These spacetimes are Petrov type-$I$, and the base metric is nonflat.
A limiting case of these spacetimes is the well-known Kasner solution, a cosmological model describing an expanding anisotropic universe, which plays an important role in generic approaches to cosmological singularities~\cite{Belinsky:1970ew,Belinski:2017awb,Garfinkle:2020lhb}.
We show that Kasner is a flat-space double copy that is not of Kerr-Schild form, and, furthermore, its Weyl curvature is of the generic form of Eq.~\eqref{twistorweyl}.
The identification of an exact, algebraically generic double copy confirms the expectations from twistor theory and is a step toward extending the double copy to a broader class of dynamical and physically interesting spacetimes.

In this work, we specialize for four dimensions, but our results can be generalized to higher dimensions. 
We reserve some technical derivations for the Appendix.

\noindent \emph{Properties of a metric double copy.}---To seek new metric double-copy examples, we first generalize beyond those studied previously. 
We assume our metric splits as
\begin{equation}
    g_{\mu \nu} = \hat{g}_{\mu \nu} + h_{\mu \nu}, 
    \label{metricsplit}
\end{equation}
where both $g_{\mu \nu}$ and the base metric $\hat{g}_{\mu \nu}$ satisfy the vacuum field equations.
Usually, $\hat g_{\mu\nu}$ is taken to be Minkowski.
In order to furnish the connection to gauge fields, we require that the test electromagnetic potential $A_\mu$ (the single copy) is derivable from $h_{\mu \nu}$.
We assume a simple relation which holds for all known examples, specifically,
\begin{align}
    A_\mu = h_{\mu\nu} \xi^\nu \,,
    \label{eq:Afromh}
\end{align}
for some vector field $\xi^\mu$ whose properties we are free to specify,
and we require $A_\mu$ satisfy vacuum Maxwell's equations on $\hat{g}_{\mu\nu}$. 
When there is a preferred choice for $\xi^\mu$, using it fixes an ambiguity in distributing functional freedom within $h_{\mu \nu}$, e.g.,~for the Kerr-Schild metric \eqref{KSDC}, between $\psi$ and $k_\mu$. 
We take the $\Phi \equiv h_{\mu \nu}\xi^\mu \xi^\nu$ as the zeroth copy, which must satisfy the wave equation on $\hat g_{\mu\nu}$. 
We note that, while we consider vacuum solutions for $g_{\mu\nu}$ and $A_\mu$, there are examples of sourced double copies \cite{Carrillo-Gonzalez:2017iyj,Easson:2021asd,Easson:2022zoh} which rely on a natural source correspondence for the Maxwell fields based upon Killing vectors (KVs).
These can be incorporated into our framework, as discussed in the Appendix.

Further constraints are needed to seek double-copy solutions.
For this, we note a key feature of the known Kerr-Schild and Weyl double copies:
The gauge potential $A_\mu$ satisfies identical Maxwell equations on both $g_{\mu\nu}$ and the base metric $\hat g_{\mu\nu}$. 
For Kerr-Schild metrics, Maxwell's equations on the base spacetime derive from Maxwell's equations on the full spacetime~\cite{Kent_2025}. 
Meanwhile, in these cases the type-$D$ Weyl double copy follows from the result that the $f_{AB}$ defined in \eqref{WDC} satisfies Maxwell's equations on the full spacetime \cite{Walker:1970un,Hughston:1972qf}.
{\it A priori}, it is not clear if this property is a necessary one for the notion of a classical double copy (and, in fact, it is not true for the dilatonic spacetimes studied in~\cite{Luna_2017,Kim:2019jwm}), but by enforcing it we identify new avenues for seeking metric double copies. 

Beginning with a field strength two-form $\boldsymbol{F}$ that satisfies the source-free Maxwell's equations with respect to the base metric $\hat g_{\mu \nu}$,
\begin{equation}
    d \boldsymbol{F} = 0, \quad d \star \boldsymbol{F} = 0, 
    \label{MaxwellDF}
\end{equation}
we may ask what metric transformations preserve $\boldsymbol{F}$ as a solution, which we call \emph{Maxwell preserving}.
It is well known that conformal transformations are of this type, but a large class of additional transformations have been identified in~\cite{Harte_2017}.
Modulo conformal rescaling, these additional transformations are additive and formed from products of the principal null directions (PNDs) of $\boldsymbol{F}$. 
In the null (degenerate) case, PNDs correspond to the direction of energy propagation.
When $\boldsymbol{F}$ is algebraically general (non-null), these can involve the two distinct real and two complex null eigenvectors of $F_{\mu \nu}$.
By insisting that $h_{\mu \nu}$ is of the form described in~\cite{Harte_2017}, we can make further progress.

In this study, we further restrict to Maxwell-preserving transformations of algebraically general field strengths involving the real PNDs $k^{(i)}_\mu = \{k_\mu, \ell_\mu\}$:
\begin{align}
   h_{\mu \nu} & = \frac{\kfactor \, k_\mu k_\nu -
   (k \cdot \ell)\kfactor\,\lfactor\, k_{(\mu}\ell_{\nu)}
   +\lfactor\,\ell_\mu \ell_\nu}
   {1-\frac{1}{4}(k \cdot \ell)^2 \kfactor\,\lfactor}\,,
   \label{eq:hnullEV}
\end{align}
where $(k \cdot \ell)$ is taken with respect to $g_{\mu \nu}$ and $\kfactor$ and $\lfactor$ being arbitrary functions of the coordinates.
Upon contraction of this with $\xi^\nu$, $A_\mu$ takes the form 
\begin{equation}
    A_\mu = \sum_{i=1}^2 V^{(i)}k^{(i)}_\mu, 
    \label{Asum}
\end{equation}
where $V^{(i)}$ are scalar functions.
However, $F_{\mu \nu}$ is itself formed from $A_\mu$, and so for each PND $k^{(j)}_\mu$ we must have
\begin{equation}
    k^{(j)}_\alpha g^{\alpha \mu} F_{\mu \nu} = \sum_{i=1}^2 2 k_{(j)}^\mu \nabla_{[\mu} \left(V^{(i)}k^{(i)}_{\nu]}\right) \propto k^{(j)}_\nu, 
    \label{the_cond}
\end{equation}
which restricts even further the form of $h_{\mu\nu}$ in Eq.~\eqref{eq:hnullEV}.

\noindent \emph{Kerr-Schild and double Kerr-Schild cases.}---Before 
introducing new metric transformations $h_{\mu\nu}$ satisfying \eqref{the_cond}, 
it is instructive to demonstrate how known classical double copies satisfy this condition. 
Consider a generalized Kerr-Schild metric \cite{TAUB1981326}
\begin{equation}
    g_{\mu \nu} = \hat{g}_{\mu \nu} + \psi k_\mu k_\nu, 
    \label{gKS}
\end{equation}
with $k_\mu$ null with respect to $g_{\mu\nu}$ and so also null with respect to $\hat g_{\mu\nu}$. 
Assuming that $A_\mu = h_{\mu \nu}\xi^\nu \equiv V k_\mu$ satisfies Maxwell's vacuum equations on $\hat{g}_{\mu \nu}$, the condition \eqref{the_cond} for $A_\mu$ to also be a solution on $g_{\mu \nu}$ corresponds to
\begin{equation}
    k_\alpha g^{\alpha \mu} F_{\mu \nu} = k^\mu \hat{\nabla}_\mu (V k_\nu) \propto k_\nu \,,
    \label{KScond}
\end{equation}
and, thus, $k_\mu$ must be geodesic.
Meanwhile, the vacuum field equations for $g_{\mu\nu}$ guarantee that $k_\mu$ is geodesic, so that these generalized Kerr-Schild spacetimes meet the double-copy conditions we have outlined and match the assumptions required for linearizing Kerr-Schild spacetimes.
In addition, for generalized Kerr-Schild metrics, if $k_\mu$ is geodesic with respect to one metric, it must be geodesic with respect to the other \cite{TAUB1981326}.
This implies this procedure works just as well mapping a Maxwell solution from $g_{\mu \nu}$ to $\hat{g}_{\mu \nu}$, as one should expect.

Another type of transformation which yields classical double-copy solutions are double-Kerr-Schild transformations \cite{Plebanski:1975wn,plebanski1976rotating,Luna:2015paa,Luna:2018dpt}, which have been analytically continued such that the expansion is with respect to two orthogonal null vectors $k_\mu \ell^\mu = 0$. 
Their metrics take the form
\begin{equation}
    g_{\mu \nu} = \hat{g}_{\mu \nu} + \psi k_\mu k_\nu + \varphi \ell_\mu \ell_\nu, \label{DKS}
\end{equation}
consistent with the form of Eq.~\eqref{eq:hnullEV}, and take both $k_\mu$ and $\ell_\mu$ to be geodesic with respect to $\hat{g}_{\mu \nu}$ and, thus, respect to $g_{\mu \nu}$. 
One can readily check that the Taub-NUT double copy \cite{Luna:2015paa} and the Pleba\'nski-Demia\'nski metric  \cite{Luna:2018dpt}, both of the form \eqref{DKS}, have potentials $A_\mu = V k_\mu + W \ell_\mu$ satisfying Eq.~\eqref{the_cond}.

\noindent \emph{Choosing $\xi^\mu$.}---To develop the final portion of our procedure, we must select a vector $\xi^\mu$ and also a procedure for identifying a test Maxwell solution.
A natural candidate for solving both problems is to use KVs, 
since it is known that for a KV $\xi^\mu$ that
\begin{equation}
    F^{\xi}_{\mu \nu} \equiv 2 \nabla_{[\mu} \xi_{\nu]}
    \label{kveqn}
\end{equation}
for $\xi_\mu \equiv g_{\mu \nu}\xi^\nu$, satisfies Maxwell's equations on Ricci-flat spacetimes~\cite{Papapetrou,Wald1974b,Wald:1984rg}.
If, in addition, $\xi^\mu$ is dual to an exact form with respect to $\hat g_{\mu\nu}$, so that $\hat g_{\mu\nu} \xi^\nu= \partial_\nu \lambda$ for some scalar $\lambda$,
then
\begin{equation}
    \xi_\mu = (\hat g_{\mu\nu} + h_{\mu\nu})\xi^\nu
    = \partial_\mu \lambda + h_{\mu \nu}\xi^\nu = A_\mu + \partial_\mu \lambda ,
    \label{eq:GaugeTransform}
\end{equation}
so that $A_\mu$ generates the same field strength $F^{(\xi)}_{\mu\nu}$.

Take $h_{\mu \nu}$ to be of the form such that Maxwell's equations are satisfied for $A_\mu$ on both the base and full metrics. 
If we additionally take $\xi^\mu$ to be a KV with respect to $\hat{g}_{\mu \nu}$, making $\xi_\mu$ covariantly constant on $\hat{g}_{\mu \nu}$, then $\Phi \equiv h_{\mu \nu}\xi^\mu \xi^\nu$ automatically satisfies the wave equation the base metric; see \eqref{zerocopyfinal}. 
Hence, we have established the conditions on when a KV may be utilized to generate a single and zeroth copy.

While the requirement of a KV that is exact with respect to $\hat g_{\mu\nu}$ appears restrictive, in fact, this connection between the KV and $A_\mu$ holds in many known examples~\cite{cdc-gem}.
This correspondence has been noted previously~\cite{Easson:2023dbk,Ortaggio1,Ortaggio2024}
and exploited to relate the Kerr-Schild and Weyl double copies~\cite{cdc-gem}. 
The condition that the KV be covariantly constant on $\hat g_{\mu\nu}$ and be gauge equivalent to $A_\mu$ conform to the ``type-$A$" categorization of Kerr-Schild double copies laid out in \cite{Bahjat-Abbas:2017htu}, which here we consider in a more general framework and so refer to as \emph{class-A} double copies.
Meanwhile those cases described as ``type-$B$" double copies are a Maxwell-preserving transformation of a class-$A$ double copy, which we describe as \emph{class-B} in our more general setting (see the Appendix for details).
One notable example of double copies that are not of class-$A$ are Petrov type-$N$ vacuum Kundt metrics, which possess a KV but whose single copies are not gauge equivalent to the Maxwell field from the KV~\cite{Ortaggio2024}.

Restricting to class-$A$ spacetimes we next utilize our construction implicit in Eqs.~\eqref{metricsplit}, \eqref{eq:Afromh}, \eqref{eq:hnullEV}, and \eqref{eq:GaugeTransform} to seek new classical double copies.

\noindent \emph{A generalized Kerr-Schild double copy.}---Given our setup,
a simple ansatz at this stage is to seek examples with a single principal null direction in the metric transformation, so that in Eq.~\eqref{eq:hnullEV} we set $\lfactor = 0$.
The result is a vacuum spacetime of generalized Kerr-Schild form, with $k^\mu$ a null geodesic and $h_{\mu\nu} = \psi k_\mu k_\nu$.
To have a KV, we focus on spacetimes which are stationary.
Specific cases of the K\'ota-Perj\'es metrics of \cite{Kota1972-gw} possess precisely these properties, as
demonstrated in \cite{Gergely_1993,Gergely_1994,Gergely_1994b}.
For these the  base metric, in this case denoted $\check{g}_{\mu \nu}$, is a type-$N$ vacuum Kundt metric.
In coordinates $\{u,v,w,\varx\}$, the line element for the base metric is
\begin{equation}
    d\check{s}^2 = -2dv ( du - 2cBw d\varx) + \frac{V_0 v}{V} \left(v^{-s}dw^2 + v^{s}d\varx^2 \right), 
    \label{baseKP}
\end{equation}
where $c\equiv \cos \eta$, $s\equiv \sin \eta$, and $\eta$, $B$, and $V_0$ are constants. 
We have defined
\begin{align}
    V \equiv V_0 \frac{v^c}{v^{2c}+B^2}.
\end{align}
The full metric is built by identifying $\psi = V$, so that~\cite{Gergely_1994}
\begin{align}
    ds^2 &= d\check{s}^2 + V (k_\mu dx^\mu)^2,
    &
    k_\mu dx^\mu &\equiv du - 2c Bwd\varx.
    \label{KPmetric}
\end{align}
It is readily apparent in these coordinates that $\partial_u = \xi^\mu \partial_\mu$ is a Killing vector on both $g_{\mu \nu}$ and $\check{g}_{\mu \nu}$. 
Moreover, $\xi^\mu$ is an exact form on $\check{g}_{\mu \nu}$ since $\check{g}_{\mu \nu}\xi^\mu dx^\nu = -dv$.
It is also easy to show that $k_\mu$ is null and geodesic, and so
our three conditions are satisfied. 
The gauge potential is
\begin{equation}
    \boldsymbol{A} = h_{\mu \nu}\xi^\nu dx^\nu = V k_\mu dx^\mu = V(du - 2cBwd\varx).
    \label{KPA}
\end{equation} 
Furthermore, since $\xi^\mu$ is Killing and exact on $\check{g}_{\mu \nu}$, this implies $\Phi \equiv h_{\mu \nu}\xi^\mu \xi^\nu$ satisfies $\check{\Box} \Phi = 0$, with $\check{\Box}$ being the wave operator on $\check{g}_{\mu \nu}$.

\noindent \emph{Kasner limit.}---There exists a physically interesting limit of Eq.~\eqref{KPmetric}, which furthermore is of primary interest due to having a flat base metric.
For $B \to 0$, which corresponds to $k_\mu$ having vanishing twist \cite{Gergely_1994c}, the base metric \eqref{baseKP} becomes a vacuum $pp$ wave which can be expressed in Kerr-Schild form. 
This means the metric can be written as
\begin{equation}
    g_{\mu \nu} = \hat{g}_{\mu \nu} 
    + \hat{\varphi}\, \hat{\ell}_\mu \hat{\ell}_\nu 
    + V k_\mu k_\nu, \label{newsplitting}
\end{equation}
where 
$\hat{\ell}_\mu$ is null with respect to $\hat{g}_{\mu \nu}$ but not $g_{\mu \nu}$, and $k_\mu$ is not null (or geodesic) with respect to $\hat{g}_{\mu \nu}$. 
Importantly, this implies $g_{\mu\nu}$ is not a double Kerr-Schild metric of the form \eqref{DKS}, but we demonstrate in the Appendix that it is of the form \eqref{eq:hnullEV}.
Note, however, that any arbitrary combination of Kerr-Schild transformations \eqref{newsplitting} can be placed into the form \eqref{eq:hnullEV} \cite{Harte:2019tid}.
The line element can be expressed as
\begin{equation}
    ds^2 = d\hat{s}^2 + \hat{\varphi} Y^2 dv^2 + V_0v^{-c}du^2, \label{GPmetric}
\end{equation}
with $\hat{\ell}_\mu dx^\mu = Y dv$, $k_\mu dx^\mu = du$, and where 
\begin{equation}
    \hat{\varphi} \equiv \frac{c s V_0^2(v^{2s}\varx^2-w^2)}{8 v^{1+c+s}} ,  
    \quad Y \equiv \frac{2}{V}.
    \quad V = V_0 v^{-c} ,
\end{equation}
Here the flat metric has the line element
\begin{equation}
    d\hat{s}^2 = -2dudv -\hat \varphi Y^2dv^2 + v^{1+c-s}dw^2 + v^{1+c+s}d\varx^2. 
    \label{flatmetric}
\end{equation}
Depending on the sign on $V_0$, the line element \eqref{GPmetric} can be transformed \cite{Gergely_1994c}(there is an error in the transformations listed in that work; we provide the corrected transformations in the Appendix.) 
into the Kasner metric or the sign-flipped Kasner metric described in \cite{McIntosh1992-ud}, and this determines if $\partial_u$ is spacelike or timelike.
These are known to be Petrov type-$I$ spacetimes, although special values of $\eta$ yield type-$D$ or flat spacetimes \cite{Stephani:2003tm}.

We have already seen that $A_\mu = V k_\mu$ is a valid vacuum gauge potential on $g_{\mu \nu}$.
We now demonstrate that Eq.~\eqref{the_cond} is satisfied, so that $A_\mu$ is a valid gauge potential for both $g_{\mu\nu}$ and the flat base metric $\hat g_{\mu\nu}$.
The condition $k_\alpha g^{\alpha \mu} F_{\mu \nu} \propto k_\nu$ 
is satisfied since $k_\mu$ is geodesic. 
By deriving the Maxwell-preserving transform from $g_{\mu\nu}$ to $\hat g_{\mu\nu}$ (from Kasner to flat), which is of the form Eq.~\eqref{eq:hnullEV}, we identify the required null vector $\ell_\mu dx^\mu \equiv du - Ydv$.
This is a PND of $F_{\mu\nu}$ on $g_{\mu\nu}$, satisfying
\begin{equation}
    \ell_\alpha g^{\alpha \mu} F_{\mu \nu} = \ell^\mu \nabla_\mu (V k_\nu) - \ell^\mu \nabla_\nu (V k_\mu) \propto \ell_\mu\,. 
    \label{newcond}
\end{equation}
Thus, we have shown that Kasner provides the first example of an exact, type-$I$, classical double copy on a flat base metric.

Consider now the Kasner line element in standard coordinates
\begin{equation}
    ds^2 = -dT^2 + T^{2p_1}dX^2 + T^{2p_2}dY^2 + T^{2p_3}dZ^2, \label{Kasner}
\end{equation}
where the powers
\begin{equation}
    p_1 \equiv \frac{1+c+s}{c+2}, \quad p_2 \equiv \frac{1+c-s}{c+2}, \quad p_3 \equiv \frac{-c}{c+2}, 
    \label{KasnerConds}
\end{equation}
satisfy the Kasner conditions $p_1 + p_2 + p_3 = 1 = p_1^2 + p_2^2 + p_3^2$ required by the field equations. 
We are interested in what the field strength $F_{\mu \nu} \equiv 2 \nabla_{[\mu}A_{\nu]}$ looks like in Cartesian coordinates.
The generalized Kerr-Schild form of Kasner requires a null direction within the Kasner universe to be chosen; by choosing it aligned in the $T$-$Z$ plane, we have introduced a preferred spatial direction. 
The Cartesian coordinates are related to the standard Kasner coordinates as $x \propto X$, $y \propto Y$, and $(t-z) \propto T^{\alpha}$ for $\alpha > 0$, and so the Kasner time $T$ becomes a null coordinate in Minkowski space. 

These distinctions generate an asymmetry in the $z$-coordinate dependence of the electric and magnetic fields.
For a stationary observer with four-velocity $u^\mu = \delta_t^\mu$, the electric field is
\begin{equation}
    \Vec{E} \propto \left(\frac{p_1 x}{(t-z)^{2+c}}, \frac{p_2 y}{(t-z)^{2+c}}, \frac{p_3}{c(t-z)^{1+c}} \right), \label{Efield}
\end{equation}
while the magnetic field is
$\Vec{B} \propto (-E_y, E_x, 0)$.
These can be shown to solve Maxwell's equations on flat spacetime.
We see that the field strength has a singularity which coincides with the Kasner singularity. 
At a fixed time, the distance to the singularity is measured by the position of an observer in the $z$ direction.
These fields represent a traveling wave with infinite density along the $t=z$ plane. The field invariants $E \cdot B = 0$ and $E^2-B^2>0$ imply there is a frame for which the field is purely electric, however, this requires an infinite boost in the $t-z$ plane.

The gauge potential $A_\mu$ is only part of the single-copy story for Kasner.
In addition, since the $pp$ wave is also a double-copy solution, there is an additional Maxwell field $\hat A = \hat \varphi (\zeta \cdot \hat l) \hat l_\mu$ on the flat and $pp$ wave spacetime, with $\zeta^\mu \partial_\mu = f(v) \partial_v$.
Solving for $f(v)$ on the Kasner spacetime yields $f(v) \propto v$.
Together, $\hat A_\mu$ and $A_\mu$ can be considered the full single-copy gauge solution for Kasner.

\noindent \emph{Kasner as a type-I Weyl double copy.}---We have now demonstrated that a type-$I$ metric double copy exists. 
However, if this algebraically generic example is an analog of the momentum space double copy, then the Weyl spinor is expected to split as in Eq.~\eqref{twistorweyl}~\cite{Chacon:2021wbr}.

A direct calculation reveals this is the case.
For Kasner, the Weyl spinor is expressed in terms of three Maxwell spinors, each related to the KVs of the spacetime.
In addition to $\partial_u$, $\partial_\varx$ and $\partial_w$ are also KVs on both $g_{\mu \nu}$ and $\hat{g}_{\mu \nu}$ but are not exact forms on the latter.
These additional KVs
provide nontrivial Maxwell solutions via Eq.~\eqref{kveqn} on both metrics, denoted $F^\varx_{\mu \nu}$ and $F^w_{\mu \nu}$, respectively.
Utilizing a tetrad for \eqref{GPmetric} aligned with $k_\mu dx^\mu = du$ and $\ell_\mu dx^\mu = du - Y dv$, the Weyl spinor can be rewritten into the form
\begin{equation}
    \Psi_{ABCD} = \frac{f_{(AB}f_{CD)}}{S} + \frac{f^\varx_{(AB}f^\varx_{CD)}}{S_\varx} + \frac{f^w_{(AB}f^w_{CD)}}{S_w} \,, \label{WeylDecomp}
\end{equation}
Here the $f_{AB}, f^\varx_{AB}$, and $f^w_{AB}$ are the spinorial analogs to $F_{\mu \nu}$, $F^\varx_{\mu \nu}$, and $F^w_{\mu \nu}$, respectively, while the scalars $S, S_\varx$, and $S_w$ are given in Eqs.~\eqref{eq:KasnerScalars}.
These satisfy the flat-space wave equation since they depend only on $v = t-z$, and, in fact, so does any scalar we extract from Weyl spinor, since by the symmetries of the spacetime $\Psi_{ABCD}$ depends only on $v$.
As such, our requirement that a zeroth copy obey the wave equation on $\hat g_{\mu\nu}$ is trivial here, although it may provide constraints in more general cases.
Nonetheless, this demonstrates for the first time that a type-$I$ Weyl double copy exists at the level of exact solutions. 

\noindent {\it Discussion}---In this Letter, we have provided a novel framework for understanding metric double copies and applied this framework to provide 
the first example of an algebraically general metric double copy on a flat base metric. 
Our framework builds on the fact that known metric double copies possess single-copy gauge potentials that solve Maxwell's equations on both the base spacetime and the full spacetime.
By focusing on a class of Maxwell-preserving metric transformations and relating the difference $h_{\mu\nu}$ between base and full metrics to the gauge potential, we have derived a set of consistency conditions which known metric double copies satisfy.
Combining these conditions with the fact that KVs provide a natural test Maxwell field, we have derived the metric double copy of the Kasner spacetime, representing an exact, algebraically general double copy.
Furthermore, the spacetime exhibits a type-$I$ Weyl double copy in accord with expectations from twistor theory \cite{Chacon:2021wbr}, opening new avenues to connect the classical and quantum double copies. 

Further extensions of this framework provide opportunities for a large number of future investigations. 
Maxwell-preserving transformations involving the complex null PNDs or for algebraically special field strengths~\cite{Harte_2017} provide a larger class of metrics to seek double copies for. 
Utilizing test Maxwell potentials that are not KVs 
could provide a path for discovering double copies in spacetimes not admitting symmetries. 
A major appeal of Kerr-Schild and certain double-Kerr-Schild spacetimes is that the index raised Ricci tensor $\tensor{R}{^\mu_\nu}$ is linear in the perturbation $h_{\mu \nu}$;
since in our framework we require that $h_{\mu \nu}\xi^\nu$ satisfies linear equations, it would be interesting to explore further restrictions such that Einstein's equations linearize in $h_{\mu \nu}$.
The fact that any 4D analytic metric may be rewritten into \emph{extended Kerr-Schild} form \cite{Ett_2010,Llosa_2009} with a flat base metric may point toward a more general formulation for a classical double copy than the framework outlined here.
Finally, higher-dimensional and nonvacuum spacetimes are not considered in this work, and both provide paths for further exploration.

\begin{acknowledgments}
We thank Elena C\'aceres, Cynthia Keeler, Ricardo Monteiro, Donal O'Connell, and Chris White for valuable conversations.
B.K. was supported by National Science Foundation (NSF) Grant No. PHY–2210562, and
A.Z. was supported by NSF Grant No. PHY-2308833. 
This paper has preprint number UT-WI-12-2025.
\end{acknowledgments}

\hspace{10mm}
 
\bibliographystyle{apsrev4-1}
\bibliography{refs.bib}

\appendix

\renewcommand{\theequation}{A\arabic{equation}}
\setcounter{equation}{0}

\section{APPENDIX}

\noindent \emph{The zeroth copy.}---Consider a metric splitting of the form $g_{\mu \nu} = \hat{g}_{\mu \nu} + h_{\mu \nu}$, and let there exist a Killing vector (KV) $\xi^\mu$ on $g_{\mu \nu}$. 
If we take this KV to also be an isometry on $\hat{g}_{\mu \nu}$, then we have
\begin{equation}
    \mathcal{L}_\xi g_{\mu \nu} = 0, \quad \mathcal{L}_\xi \hat{g}_{\mu \nu} = 0 \quad \implies \quad \mathcal{L}_\xi h_{\mu \nu} = 0 \,.
\end{equation}
As established in the main text, we require $\xi^\mu$ to be an exact form on $\hat{g}_{\mu \nu}$ to ensure $\xi_\mu$ is gauge equivalent to $A_\mu = h_{\mu \nu}\xi^\nu$. 
By being both exact and Killing on $\hat{g}_{\mu \nu}$, $\xi^\mu$ must be covariantly constant on $\hat{g}_{\mu \nu}$, meaning that the Lie derivative in terms of the metric-compatible $\hat{\nabla}_\mu$ can be expressed as
\begin{equation}
    \mathcal{L}_\xi h_{\mu \nu} = 0 = \xi^\alpha \hat{\nabla}_\alpha h_{\mu \nu}\,. \label{LieDerivh}
\end{equation}
If we restrict $h_{\mu \nu}$ to be  Maxwell-preserving transformation on $\hat g_{\mu\nu}$ 
then Maxwell's equations for $A_\mu$ become
\begin{equation}
    \hat{\nabla}^\nu \left[ \hat{\nabla}_\mu (h_{\nu \alpha}\xi^\alpha) - \hat{\nabla}_\nu (h_{\mu \alpha}\xi^\alpha) \right] = 0\,.
\end{equation}
By contracting in $\xi^\mu$, we find the first term vanishes by \eqref{LieDerivh}, with the resulting equation becoming
\begin{equation}
    \hat{\Box}\Phi \equiv \hat{\Box}(h_{\mu \nu}\xi^\mu \xi^\nu) = 0\,.\label{zerocopyfinal}
\end{equation}
Hence, these sets of assumptions ensure a zeroth copy exists.

\noindent \emph{Killing sources.}---The sourced classical double-copy prescriptions utilized in \cite{Carrillo-Gonzalez:2017iyj,Easson:2021asd,Easson:2022zoh} rely upon the same source prescription as those of a two-form constructed from a KV, $\xi^\mu$ which satisfies
\begin{equation}
    \nabla_\nu F^{\mu \nu} = 2 \tensor{R}{^\mu_\nu}\xi^\nu,
    \label{eq:SourcedMaxwell}
\end{equation}
for $F_{\mu \nu} \equiv 2 \nabla_{[\mu}\xi_{\nu]}$. 
The single copies on the flat or base spacetimes considered in \cite{Carrillo-Gonzalez:2017iyj,Easson:2021asd,Easson:2022zoh} 
have the same source prescription as Eq.~\eqref{eq:SourcedMaxwell} and, hence, always make reference to the original spacetime's Ricci tensor and use the full metric to raise indices on the source. 
As such, our Maxwell-preserving transformations which preserve the left-hand side of
\begin{equation}
    d \star \boldsymbol{F} = \star \boldsymbol{J}
\end{equation}
are applicable in these cases, as long as the right-hand side always makes reference to the full spacetime metric which constructs $R_{\mu \nu}$.

\noindent \emph{``Class-B" spacetimes.}---Here, we explain how \emph{type-B} double copies as defined in Ref.~\cite{Bahjat-Abbas:2017htu} fall into our framework.
Consider a \emph{class-A} double copy as defined in the main body, for simplicity as a generalized Kerr-Schild metric as in Eq.~\eqref{gKS}, meaning
$A_\mu = \kfactor (k \cdot \xi) k_\mu$ is a vacuum Maxwell potential on both $g_{\mu \nu}$ and $\hat{g}_{\mu \nu}$. 
However, by construction, a new Maxwell-preserving transformation $\hat{h}_{\mu \nu}$ of the base metric ensures the potential $A_\mu$ is also a solution on $\hat{g}_{\mu \nu} \to \hat{g}_{\mu\nu} + \hat{h}_{\mu \nu}$.
 
In Ref.~\cite{Bahjat-Abbas:2017htu}, the authors demonstrated how Kerr-Schild double copies can be split between $\hat{g}_{\mu \nu}$ and $h_{\mu \nu}$; e.g.,~the vacuum Schwarzschild single copy was also a valid single copy on (A)dS space, which can themselves be written in Kerr-Schild form. 
However, the result is actually stronger: A given single copy is valid on an \textit{arbitrary} spacetime related to the original via any Maxwell-preserving transformation. 
Note that this necessarily changes the zeroth-copy prescription, as also noted in \cite{Bahjat-Abbas:2017htu}.

A concrete example which implicitly utilizes this correspondence beyond Kerr-Schild form are the Kerr-Taub-NUT and vacuum Pleba\'nski-Demia\'nski double copies analyzed in Ref.~\cite{Luna:2018dpt}. The Kerr-Taub-NUT spacetime contains a Killing-Yano tensor and so has a KV gauge equivalent to its single copy as outlined in \cite{cdc-gem}, making it a class-$A$ double copy. 
As most easily seen via the coordinates utilized in \cite{Luna:2018dpt}, the Kerr-Taub-NUT metric $\tilde{g}_{\mu \nu}$ is related to the vacuum Pleba\'nski-Demia\'nski metric via a Maxwell-preserving transformation
\begin{equation}
    \tilde{g}_{\mu \nu} \to \Omega^2\left( \tilde{g}_{\mu \nu} + \kfactor k_\mu k_\nu + \lfactor \ell_\mu \ell_\nu \right),
\end{equation}
where $\kfactor,\lfactor$, and $\Omega$ are functions of coordinates and $k_\mu$ and $\ell_\mu$ are PNDs of the single-copy field strength $F_{\mu \nu}$ satisfying $k_\mu \ell^\mu = 0$. 
This explains why the single copy of the vacuum Pleba\'nski-Demia\'nski metric utilized in \cite{Luna:2018dpt} is identical to the Kerr-Taub-NUT single copy.
That they exhibit different zeroth copies is not surprising, and the relation for any type-$D$ vacuum spacetime
\begin{equation}
    \Box (\Psi_2)^{1/3} \equiv \Box S = \Psi_2 S
\end{equation}
ensures that if $\Psi_2 \propto a$ for some constant $a$, then $a^{-1/3}S$ satisfies a wave equation on the flat spacetime defined by the limit $a \to 0$. 

\noindent \emph{Kasner as a Maxwell-preserving transformation}---In order to demonstrate that the metric \eqref{newsplitting} can be placed into the form of Eq.~\eqref{eq:hnullEV}, specifically a Maxwell-preserving transformation of flat space, note that our field strength $F_{\mu \nu}$ has different eigenvectors on $g_{\mu \nu}$ and $\hat{g}_{\mu \nu}$. 
Hence, the difference of the metrics $h_{\mu \nu}=g_{\mu \nu}-\hat{g}_{\mu \nu}$ can be expressed in two different forms of Eq.~\eqref{eq:hnullEV}, depending on which eigenvectors we express the transformation in terms of.
In other words, we can use $h_{\mu\nu}$ to transform from the flat $\hat{g}_{\mu\nu}$ to $g_{\mu\nu}$, where it is most naturally expressed in terms of PNDs on $\hat{g}_{\mu\nu}$, or $-h_{\mu\nu}$ to transform from $g_{\mu\nu}$ to $\hat{g}_{\mu\nu}$, where it is natural to express in terms of PNDs on $g_{\mu\nu}$. 
It is the latter that we show first.

Beginning with a Maxwell-preserving transformation from $g_{\mu \nu}$ to $\hat{g}_{\mu \nu}$, we note that $k_\mu$ and
\begin{equation}
    \ell_\mu dx^\mu 
    \equiv du - Ydv   
    = k_\mu dx^\mu - \hat{\ell}_\mu dx^\mu \label{elldef}
\end{equation}
are the real PNDs of $F_{\mu \nu}$ on $g_{\mu \nu}$. 
It is straightforward to show
\begin{equation}  
    \ell_\alpha g^{\alpha \mu} \nabla_\mu (V k_\nu) = 0, \quad \ell_\alpha g^{\alpha \mu} \nabla_\nu (V k_\mu) = \frac{V c}{v}\ell_\nu,
    \label{futureref}
\end{equation}
which are the self-consistency conditions \eqref{the_cond}. 
Furthermore utilizing $k_\mu$ and $\ell_\mu$ as the PNDs in Eq.~\eqref{eq:hnullEV} (used here to construct $\hat h_{\mu\nu}$), along with
\begin{equation}
    k \cdot \ell = \frac{2}{V}, \quad \kfactor = V, \quad \lfactor = \frac{V\hat{\varphi}}{V+\hat{\varphi}},
\end{equation}
it is direct to show that
\begin{equation}
    g_{\mu \nu} = \hat{g}_{\mu \nu} + h_{\mu \nu} = \hat{g}_{\mu \nu} + \hat{\varphi}\hat{\ell}_\mu \hat{\ell}_\nu + V k_\mu k_\nu
\end{equation}
verifying that 
$h_{\mu \nu}$ is a Maxwell-preserving transform.

Alternatively, transforming from $\hat{g}_{\mu \nu}$ to $g_{\mu \nu}$ requires the real PNDs of $F_{\mu \nu}$ on $\hat{g}_{\mu \nu}$, for which we define $\hat{h}_{\mu \nu}$ which has the same functional form as \eqref{eq:hnullEV} except the PNDs in the expressions are $k_\mu \to \hat{k}_\mu$ and $\ell_\mu \to \hat{\ell}_\mu$. The PNDs themselves are $\hat{\ell}_\mu$ as previously defined and
\begin{equation}
    \hat{k}_\mu \equiv \hat{\varphi}\hat{\ell}_\mu + V k_\mu, 
    \quad  {\textrm{with}} \qquad
    \hat{k} \cdot \hat{\ell} = -2.
\end{equation}
Utilizing $\hat{k}_\mu$ and $\hat{\ell}_\mu$ as the PNDs in Eq.~\eqref{eq:hnullEV}, as well as the substitutions
\begin{align}
    \lfactor &= -\hat{\varphi}, & \kfactor &= -(V+\hat{\varphi})^{-1},
\end{align}
yields
\begin{equation}
    \hat{g}_{\mu \nu} = g_{\mu \nu} + \hat{h}_{\mu \nu} = g_{\mu \nu} - \hat{\varphi}\hat{\ell}_\mu \hat{\ell}_\nu - V k_\mu k_\nu.
\end{equation}

\noindent \emph{Details of the Kasner single copy}---Consider now the Kasner metric
\begin{equation}
    ds^2 = -dT^2 + T^{2p_1}dX^2 + T^{2p_2}dY^2 + T^{2p_3}dZ^2.
\end{equation}
We can transform to the metric \eqref{GPmetric} by utilizing the coordinate transformation
\begin{equation}
    \begin{aligned}
        T &= \frac{1}{\delta \sqrt{V_0}}v^\delta,
        \quad X=(\delta \sqrt{V_0})^{p_1} \varx,
        \quad Y=(\delta \sqrt{V_0})^{p_2} w,\\
        Z&=(\delta \sqrt{V_0})^{p_3}\left[u\sqrt{V_0} - \frac{v^{1+c}}{(1+c)\sqrt{V_0}} \right], \label{KastoGP}
    \end{aligned}
\end{equation}
where $\delta \equiv \frac{1}{2}(c + 2)$, and $V_0 > 0$. 
The case $V_0 < 0$ in Eq.~\eqref{GPmetric} corresponds to the sign-flipped Kasner metric \cite{McIntosh1992-ud} with line element
\begin{equation}
    ds^2 = dT^2 + T^{2p_1}dX^2 + T^{2p_2}dY^2 -T^{2p_3}dZ^2,
\end{equation}
and the coordinate transformation from this metric to that of \eqref{GPmetric} is given by $V_0 \to -V_0$ in \eqref{KastoGP}, along with a change in relative sign within $Z$. 
We focus on the Kasner metric when discussing its single copy.

Note that we have utilized a specific choice of parametrization of the Kasner conditions \eqref{KasnerConds} which will manifest itself within our single copy but could have chosen any spatial coordinate $\{X,Y,Z\}$ to have $u$ dependence. 
Because of the coordinate singularity when $V_0 \to 0$, we cannot utilize Kasner coordinates to understand the behavior of the Maxwell field on the base metric. 

Since Maxwell's equations in vacuum are scale invariant, we write matrix expressions only up to proportionality. 
In the coordinates used in Eq.~\eqref{GPmetric} we have $T \propto v^\delta$, and so $F_{\mu \nu} \propto V_0 v^{-1-c} \delta^u_{[\mu}\delta^\varx_{\nu]}$.
Note that this has the same form in the flat metric given in Eq.~\eqref{flatmetric}; however, this form is difficult to extract electric and magnetic fields from. 
We can utilize the coordinate transformation
\begin{equation}
    \begin{aligned}
        u &= \frac{1}{\sqrt{2}}(t+z) - \frac{c+2}{4v}\left[p_1x^2 + p_2y^2 \right],\\
        v &= \frac{1}{\sqrt{2}}(t-z),
        \quad \varx = \frac{x}{\sqrt{v^{1+c+s}}},
        \quad w = \frac{y}{\sqrt{v^{1+c-s}}},
    \end{aligned}
\end{equation}
to arrive at Cartesian Minkowski coordinates, for which the electric field for the observer $u^\mu = \delta^\mu_t$ becomes \eqref{Efield}.

In order to ascertain the spinorial form for a Weyl double copy, one can utilize a null tetrad for the metric \eqref{GPmetric},
\begin{equation}
    \begin{aligned}
        \tilde{k}_\mu dx^\mu &= v^{-c/2}\sqrt{\frac{V_0}{2}}du,\\
        \tilde{\ell}_\mu dx^\mu &= -v^{-c/2}\sqrt{\frac{V_0}{2}}du + v^{c/2}\sqrt{\frac{2}{V_0}}dv,\\
        m_\mu dx^\mu &= -\sqrt{\frac{v^{1+c+s}}{2}}d\varx + i\sqrt{\frac{v^{1+c-s}}{2}}dw,\\
    \end{aligned}
\end{equation}
where the tildes over $\tilde{k}_\mu$ and $\tilde{\ell}_\mu$ are used to distinguish them from the one-forms $k_\mu$ and $\ell_\mu$ utilized in the main text of this work. 
From the null tetrad, one can construct a vierbein and, hence, soldering symbols $\tensor{\sigma}{^\mu_{A \dot{A}}}$ from the tangent space into spin space.
Any type-$I$ Weyl spinor can be rewritten (for a normalized spin frame $o_A \iota^A = 1$) into the form \cite{Stephani:2003tm}
\begin{equation}
\begin{aligned}
    \Psi_{ABCD} = &\Psi_0 o_{A}o_Bo_Co_{D} + \Psi_4 \iota_{A}\iota_B\iota_C\iota_{D}\\
    &+ 6\Psi_2 o_{(A}\iota_Bo_C\iota_{D)}, \label{typeIweyl}
\end{aligned}
\end{equation}
where $\{o_A, \iota_A\}$ form a normalized spinor basis $o_A \iota^A = 1$, with $\tilde{k}^\mu = \sigma^\mu_{A \dot{A}} o^A o^{\dot{A}}$ and $\tilde{\ell}^\mu = \sigma^\mu_{A \dot{A}} \iota^A \iota^{\dot{A}}$, and $\Psi_i$ are the Weyl scalars~\cite{newman1962approach},
where here $\Psi_0 = \Psi_4$. 
The Weyl scalars for this tetrad are
\begin{equation}
    \Psi_0 = \Psi_4 = -\frac{V  c  s}{4v^2} , \quad \Psi_2 = -\frac{V(1+c)c}{4v^2} \,,
\end{equation}
while the spinor analog to $F_{\mu \nu}$ can be shown to be
\begin{equation}
    F_{\mu \nu} \to f_{AB} = -V\, c v^{-1}o_{(A}\iota_{B)} \,.
\end{equation}
Meanwhile,
\begin{equation}
    F^\pm_{\mu \nu} \to f^{\pm}_{AB} \equiv \frac{1}{2}(1+c \pm s)\sqrt{v^{s-1} V_0}\left(\iota_A \iota_B \mp o_A o_B \right),
\end{equation}
for $F^+_{\mu \nu} \equiv F^\varx_{\mu \nu}$ and $F^-_{\mu \nu} \equiv F^w_{\mu \nu}$, so upon the identification
\begin{align}
\label{eq:KasnerScalars}
    S & \equiv -2v^{-c}V_0 c[3(1+c)]^{-1}, \\
    S_\varx & \equiv -2v^{1+c+s}(1+c+s)^2 (cs)^{-1},\\
    S_w & \equiv -4v^{1+c-s} (1+c)(s-1) (cs)^{-1},
\end{align}
the Weyl spinor can be rewritten into the form \eqref{WeylDecomp}.

\end{document}